\documentclass[preprint,aps,nofootinbib]{revtex4-1}
\pdfoutput=1

\usepackage{amsmath}  
\usepackage{amsbsy,amssymb,amsfonts}
\usepackage{graphicx}
\usepackage{color}
\usepackage{epsf}
\usepackage[utf8x]{inputenc}
\usepackage{hyperref}
\usepackage{multirow}
\def\ie{{\em{i.e.}},}

\def\bravert{\egroup\,\vrule\,\bgroup}

{\catcode`\|=\active
  \gdef\Twoint#1{\left(\mathcode`\|"8000\let|\bravert {#1}\right)}}
{\catcode`\|=\active
  \gdef\Braket#1{\left<\mathcode`\|"8000\let|\bravert {#1}\right>}}

\newcommand{\beq}{\begin{equation}}
\newcommand{\eeq}{\end{equation}}
\newcommand{\beqa}{\begin{eqnarray}}
\newcommand{\eeqa}{\end{eqnarray}}
\newcommand{\bea}{\begin{array}}
\newcommand{\eea}{\end{array}}

\newcommand{\bef}{\begin{figure}}
\newcommand{\ef}{\end{figure}}
\newcommand{\bc}{\begin{center}}
\newcommand{\ec}{\end{center}}
\newcommand{\bt}{\begin{table}}
\newcommand{\et}{\end{table}}
\newcommand{\btb}{\begin{tabular}}
\newcommand{\etb}{\end{tabular}}

\newcommand{\au}{{\em a.u.}}

\def\rvac{\left| \rule{0.3cm}{.0cm} \right>}

\def\au{{\it a.u.\ }}
\begin{document}

\title { Variational Calculation of the Hyperfine Stark Effect in Atomic $^{87}$Rb, $^{133}$Cs, and $^{169}$Tm }

\vspace*{1cm}

\author{Timo Fleig}
\email{timo.fleig@irsamc.ups-tlse.fr}
\affiliation{Laboratoire de Chimie et Physique Quantiques,
             FeRMI, Universit{\'e} de Toulouse,
             118 Route de Narbonne, 
             F-31062 Toulouse, France }

\vspace*{1cm}
\date{\today}

\vspace*{1cm}
\begin{abstract}
An electronically variational approach to the calculation of atomic hyperfine structure transition energies under the 
influence of static external electric fields is presented.
The method avoids the calculation of intermediate atomic states entirely and requires only the wavefunctions of the
electronic states involved in the respective hyperfine levels.
These wavefunctions are obtained through relativistic general-excitation-rank configuration interaction theory.
A variant of the method also enables for calculations on atoms with the most complicated of shell structures.

Applications to $^{87}$Rb, $^{133}$Cs and a specific clock transition in $^{169}$Tm are presented. The final results
$k_{\text{Rb}} = -1.234 \pm 0.0223$ [$10^{-10}$ Hz/((V/m)$^2$)] and $k_{\text{Cs}} = -2.347 \pm 0.084$ [$10^{-10}$ Hz/((V/m)$^2$)] 
obtained under inclusion of up to quintuple excitations in the atomic wavefunction expansion
are compatible with previous calculations and, in the case of Cs, confirm that one of the earlier experimental measurements
is not reliable.
For $^{169}$Tm that is used in the development of atomic clocks the differential static scalar electric
dipole polarizability between ground levels $J=\frac{7}{2}$ and $J=\frac{5}{2}$ is calculated to be
$\Delta\alpha^s_0 = -0.134 \pm 0.11$ a.u.  This result from a pure {\it{ab initio}} calculation confirms
the result of $\Delta\alpha^s_0 = -0.063^{+0.01}_{-0.005}$ a.u.  obtained in {\it{Nat. Comm.}} {\bf{10}} (2019) 1724 where 
a combination of measurement and theoretical modeling has been used.
\end{abstract}

\maketitle
\section{Introduction}
\label{SEC:INTRO}
Among the various effects \cite{ludlow_rmp_2015} contributing to the frequency uncertainty of an atomic clock the blackbody 
radiation (BBR) shift plays a major role \cite{Itano_Wineland_Blackbody1982,porsev_derevianko_PRA2006}. The BBR shift, a
systematic environmental perturbation, can be 
approximately related to the coefficient of the static electric-field shift of the clock transition 
\cite{Angstmann_Dzuba_Flambaum_PRL2006,Kozlov_Dzuba_Flambaum_PRA2014}. The latter coefficient, in turn, describes the
difference between the static electric dipole polarizabilities of the respective clock-transition states 
\cite{Safronova_BBR_2009}.
Their calculation typically requires a summation over a complete (bound and scattering) set of intermediate atomic states
which can be problematic and has in the past led to disagreements (\cite{Angstmann_Dzuba_Flambaum_PRA2006} and references
therein). 

If the clock transition involves atomic hyperfine states then the static electric-field shift of those hyperfine levels in
the external electric field \cite{schwartz_SPT_1959,Fortson_Ramsey_1964} needs to be determined.
The present approach takes an external electric field into account variationally -- {\em{i.e.}}, to infinite order expressed
in the language of perturbation theory -- in the optimization of the atomic wavefunction. In this way, the summation over 
intermediate states \cite{Angstmann_Dzuba_Flambaum_PRA2006,beloy_derevianko_PRL2006,Safronova_UI_MS_2009} is avoided 
and only the field-dependent target-state 
electronic wavefunction of the atom has to be determined. The static Stark shift of a hyperfine clock transition then results
from a simple expectation value of this wavefunction over the magnetic hyperfine operator.

Difficulties in addressing atomic quantum states with several
open electronic shells ({\em{i.e.}}, containing several unpaired electrons) and/or atomic states with a few holes in otherwise
filled shells have been reported \cite{Safronova_BBR_2009}. The present approach and method has no such particular difficulties
and the wavefunctions for more complicated quantum states of atoms can be obtained straightforwardly. This has recently been
demonstrated in the calculation of electronic electric quadrupole moments of the thulium atom \cite{Fleig_PhysRevA.107.032816}
which comprises an open $f$ shell in its electronic ground state.

Indeed, hyperfine-level transitions in the thulium atom have recently been used as clock transitions in the design of an optical
atomic clock with unusually low sensitivity to the BBR \cite{NatComm_Tm_2019,PRA_Tm_2016,PRA_Tm_2020}. The relevant BBR 
frequency shift has been experimentally estimated \cite{NatComm_Tm_2019} to be a few orders of magnitude smaller than the 
corresponding shift in clock transitions of other neutral atoms. The uncertainty of this measurement has been given as
around $50$\% \cite{AIP_Tm_2020} for the electric component of the BBR. An extensive theoretical study of the thulium
atom with relevance to its use as an atomic clock has been presented very recently \cite{fritzsche_Tm_2024}, but the BBR
or hyperfine Stark shifts have not been addressed in that work. 

The paper is organized as follows. In section \ref{SEC:THEORY} the theoretical approach is laid out in detail and a comparison
with the common perturbative approaches to the calculation of hyperfine Stark shifts is drawn. In the following section 
\ref{SEC:RESULTS}
the hyperfine Stark coefficients for three atoms of relevance to atomic-clock research are presented using the present
methods.
In addition, the differential static dipole polarizability relevant to the clock transition in $^{169}$Tm is calculated using
a finite-field approach that has earlier been applied to the calculation of spin-orbit resolved electric dipole
polarizabilities in atomic states \cite{fleig_halo,fleig_pol13}. 
In the final section \ref{SEC:CONCL} conclusions from the presented work are drawn.    
\section{Theory}
\label{SEC:THEORY}

The present approach differs substantially from previous approaches 
\cite{Angstmann_Dzuba_Flambaum_PRA2006,Safronova_BBR_2009,beloy_derevianko_PRL2006} 
to calculate electric polarizabilities of atoms including the hyperfine interaction. Before presenting the formal details it is, 
therefore, in place to lay out the big picture of the present method.

In a hypothetical atom without relativistic effects -- in particular without the spin-orbit interaction -- polarizabilities can be defined
for orbital angular momentum ($L$) microstates and denoted $\alpha_{L,M_L}$. If now very weak spin-orbit interaction is
taken into account then the total angular momentum $J$ becomes an exact quantum number. Under the assumption, however, that
the magnetic coupling is very weak, $M_L$ remains an approximately valid quantum number and the corresponding polarizabilities
can be labelled $\alpha_{J,L,(M_L)}$. For clarity of notation, the approximate quantum number is set in parentheses.

By analogy, polarizabilities for atoms without magnetic hyperfine interaction but including the electronic spin-orbit
interaction
can be defined exactly as $\alpha_{J,M_J}$. Including now a very weak magnetic hyperfine interaction leaves $M_J$ as an
approximately valid quantum number and the polarizabilities are correspondingly labelled $\alpha_{F,J,(M_J)}$ where 
$F \in \{J+I, \ldots, |J-I|\}$ and $I$ is the nuclear spin quantum number. This is the picture adopted in the present
approach.

In terms of methodology,
the first step consists in variationally determining the wavefunctions of the atomic electronic states in question including
relativistic effects, electron correlation effects, and fully relaxing the wavefunction with respect to the externally
applied finite uniform electric field. 
In the second step, the obtained wavefunctions are used as zeroth-order wavefunctions in a perturbative first-order evaluation
of the hyperfine interaction in the relevant atomic states.

\subsection{Static Hyperfine Stark Shift}

The following formulation is energy-based and in its present form only applicable to atomic $S$ (orbital angular momentum free)
states where non-scalar static electric polarizability is zero. The more general case will be presented in forthcoming
work. Thus, the present formulation is applied to Rubidium and Cesium electronic ground states. For Thulium, a different
approach is chosen as to be discussed below.

We suppose a hyperfine doublet of states denoted with total angular-momentum quantum numbers $F_u$ (upper) and 
$F_l$ (lower) is subjected to a static external electric field $E_{\text{ext}}$. Treating the hyperfine interaction
to first order in perturbation theory, the associated field-dependent transition energy can be written as
\begin{equation}
 \Delta\varepsilon(E_{\text{ext}}) = \varepsilon_{J_u,({M_J}_u)}(E_{\text{ext}}) + \varepsilon_{F_u}(E_{\text{ext}})
                            - \big( \varepsilon_{J_l,({M_J}_l)}(E_{\text{ext}}) + \varepsilon_{F_l}(E_{\text{ext}}) \big).
 \label{EQ:FDTE}
\end{equation}
where $J$ denotes total electronic angular momentum associated with the respective hyperfine level. 
$\varepsilon_{F}$ is thus the hyperfine energy relative to the
respective electronic reference energy $\varepsilon_{J,M_J}$.

This transition energy can be related to a transition frequency which is also field dependent:
\begin{equation}
 \nu(E_{\text{ext}}) = \frac{\Delta\varepsilon(E_{\text{ext}})}{h}
\end{equation}
where $h$ is Planck's constant. The shift of this transition frequency due to $E_{\text{ext}}$ is in the present evaluated
by using a finite but small $E$ field and a zero-field calculation. Then
\begin{equation}
 \delta\nu(E_{\text{ext}}) = \nu(E_{\text{ext}}) - \nu(E_{\text{ext}} = 0).
\end{equation}
Following the definition in Ref. \cite{Safronova_BBR_2009} with according modifications and taking the electronic
degrees of freedom into account the Stark coefficient $k$ as a function of this frequency shift
is expressed as
\begin{eqnarray}
 \nonumber
 k = \frac{\delta\nu(E_{\text{ext}})}{E_{\text{ext}}^2} 
      &=& \frac{1}{hE_{\text{ext}}^2}
 \left[\big(\varepsilon_{J_u,({M_J}_u)}(E_{\text{ext}}) + \varepsilon_{F_u}(E_{\text{ext}})
                 - \varepsilon_{J_l,({M_J}_l)}(E_{\text{ext}}) - \varepsilon_{F_l}(E_{\text{ext}})\big) \right. \\
 \label{EQ:K_DEF}
    && \hspace{1.5cm} \left.  -\big(\varepsilon_{J_u,({M_J}_u)}(0) + \varepsilon_{F_u}(0)
                 - \varepsilon_{J_l,({M_J}_l)}(0) - \varepsilon_{F_l}(0)\big)\right]  \\
 \nonumber
      &=& \frac{1}{hE_{\text{ext}}^2}
 \left[\big(\varepsilon_{J_u,({M_J}_u)}(E_{\text{ext}}) - \varepsilon_{J_l,({M_J}_l)}(E_{\text{ext}})\big)
                   + \big(\varepsilon_{F_u}(E_{\text{ext}}) - \varepsilon_{F_l}(E_{\text{ext}})\big) \right. \\
    && \hspace{1.5cm} \left.  -\big(\varepsilon_{J_u,({M_J}_u)}(0) - \varepsilon_{J_l,({M_J}_l)}(0)\big)
                  - \big(\varepsilon_{F_u}(0) - \varepsilon_{F_l}(0)\big)\right] 
 \label{EQ:K_DEF2}
\end{eqnarray}
and can, therefore, be calculated if the field-dependent level energies are known. 

At this point a distinction between two cases can be made:
\begin{enumerate}
 \item The two respective hyperfine levels belong to different electronic states. Then the
       Stark coefficient is stongly dominated by the electric polarizability difference between the different
       electronic wavefunctions and it can be written as
       \begin{equation}
        k \approx \frac{1}{hE_{\text{ext}}^2}
           \left[ \big(\varepsilon_{J_u,({M_J}_u)}(E_{\text{ext}}) - \varepsilon_{J_l,({M_J}_l)}(E_{\text{ext}})\big)
                - \big(\varepsilon_{J_u,({M_J}_u)}(0) - \varepsilon_{J_l,({M_J}_l)}(0)\big) \right]
        \label{EQ:K_EL}
       \end{equation}
 \item The two respective hyperfine levels belong to the same electronic state. In this case all
       of the electronic energies in the expression (\ref{EQ:K_DEF2}) cancel pairwise and the Stark coefficient is 
       \begin{equation}
        k = \frac{1}{hE_{\text{ext}}^2}
            \left[ \big(\varepsilon_{F_u}(E_{\text{ext}}) - \varepsilon_{F_l}(E_{\text{ext}})\big)
                 - \big(\varepsilon_{F_u}(0) - \varepsilon_{F_l}(0)\big)\right]
        \label{EQ:K_HF}
       \end{equation}
\end{enumerate}

\subsection{Field-Dependent Hyperfine Level Energy}

We will now be concerned with the above second case.
A field-dependent hyperfine energy is related to the magnetic hyperfine constant $A$ as (see \cite{bethe_hypfin}, p. 110)
\begin{equation}
 \varepsilon_F(E_{\text{ext}}) \approx \frac{1}{2}\, \left[ F(F+1) - I(I+1) - J(J+1) \right]\, A(E_{\text{ext}})
 \label{EQ:CG_E}
\end{equation}
where the quantum numbers $I$ and $J$ refer to nuclear spin and to total electronic angular momentum, respectively, of 
the state in question. The relationship Eq. (\ref{EQ:CG_E}) is not exact because the presence of the external field
lifts the full rotational symmetry of the atom and thus also the strict validity of total angular momentum as a good
quantum number. However, since the applied fields are very small (see below) the relation is still approximately correct.

Thus, if the field-dependent hyperfine constant $A(E_{\text{ext}})$ can be determined then also the Stark coefficient $k$.
Following the implementation in Refs. \cite{Fleig2014,Fleig_Skripnikov2020} $A(E_{\text{ext}})$ is calculated as an 
expectation value
\begin{equation}
 A(E_{\text{ext}}) = 
 \left< \hat{H}_{\text{HF}} \right>_{\psi(E_{\text{ext}})}
 \label{EQ:A_CONST}
\end{equation}
with the one-body hyperfine Hamiltonian
\begin{equation}
 \hat{H}_{\text{HF}} = -\frac{\mu[\mu_N]}{2cIm_p M_J}\,
 \sum\limits_{i=1}^n\, \left( \frac{\boldsymbol{\alpha}_i \times {\bf{r}}_{i}}{r_{i}^3} \right)_z
 \label{EQ:HHF}
\end{equation}
over a field-dependent electronic wavefunction $\psi(E_{\text{ext}})$ where $n$ is the number of electrons,
$\boldsymbol{\alpha}$ is a Dirac matrix, 
$\mu$ is the nuclear magnetic moment [in nuclear magnetons], $\frac{1}{2cm_p}$ is the nuclear magneton in \au, 
$m_p$ is the proton rest mass, and ${\bf{r}}$ is the electron position operator.

$\psi(E_{\text{ext}})$ is obtained by solving
\begin{equation}
 \hat{H}({{E}_{\text{ext}}}) \big| \psi(E_{\text{ext}}) \big>
        = \varepsilon({{E}_{\text{ext}}}) \big| \psi(E_{\text{ext}}) \big>
 \label{EQ:PSI_VAR}
\end{equation}
with $\varepsilon({{E}_{\text{ext}}})$ the field-dependent energy eigenvalue and $\hat{H}({{E}_{\text{ext}}})$ the
Dirac-Coulomb (DC) Hamiltonian including the interaction term with the external field:
\begin{eqnarray}
 \nonumber
 \hat{H}({{E}_{\text{ext}}}) &:=& \hat{H}^{\text{Dirac-Coulomb}} + \hat{H}^{\text{Int-Dipole}} \\
  &=& \sum\limits^n_j\, \left[ c\, \boldsymbol{\alpha}_j \cdot {\bf{p}}_j + \beta_j c^2
-
  \frac{Z}{r_{jK}}{1\!\!1}_4 \right]
+ \sum\limits^n_{k>j}\, \frac{1}{r_{jk}}{1\!\!1}_4
  + \sum\limits_j\, {\bf{r}}_j \cdot {\bf{E}_{\text{ext}}}\, {1\!\!1}_4
 \label{EQ:HAMILTONIAN}
\end{eqnarray}
${\bf{E}_{\text{ext}}} = E_z {\bf{e}}_z$ is uniform in space, the indices $j,k$ run over $n$ electrons, $Z$ is the proton
number with the nucleus $K$ placed at the origin, and $\boldsymbol{\alpha}, \beta$ are standard Dirac matrices.
$E_z$ is not treated as a perturbation but included {\it{a priori}} in the variational
optimization \cite{DIRAC_JCP} of the wavefunction, $\psi(E_z)$.

Technically, $\psi(E_z)$ is a configuration interaction (CI) vector \cite{knecht_luciparII} built from
Slater determinants over field-dependent 4-spinors. This vector is represented as
\begin{equation}
        \left| \alpha\, J\, M_J \right> \equiv \sum\limits_{K=1}^{\rm{dim}{\cal{F}}^t(M,n)}\,
                                       c_{(\alpha, J, M_J),K}\, ({\cal{S}}{\overline{\cal{T}}})_K \rvac
        \label{EQ:AT_WF}
\end{equation}
where ${\cal{F}}^t(M,n)$ is the symmetry-restricted sector of Fock space with $n$ electrons in $M$ four-spinors,
${\cal{S}} = a^{\dagger}_i a^{\dagger}_j a^{\dagger}_k \ldots$ is a string of spinor creation operators,
${\overline{\cal{T}}} = a^{\dagger}_{\overline l} a^{\dagger}_{\overline m} a^{\dagger}_{\overline n} \ldots$
is a string of creation operators of time-reversal transformed spinors. The determinant expansion coefficients
$c_{(\alpha, J, M_J),K}$ are generally obtained as described in refs. \cite{fleig_gasci,fleig_gasci2}
by diagonalizing the Dirac-Coulomb Hamiltonian.

\subsection{Comparison with Perturbation Theory}

In order to compare the present method with the perturbative approaches typically used in the literature, it is
assumed that a zeroth-order CI problem for the target state $\psi_0$ is solved where $\hat{V} = \hat{H}^{\text{Int-Dipole}} = 
\sum\limits_j\, {\bf{r}}_j \cdot {\bf{E}_{\text{ext}}}\, {1\!\!1}_4$ is not included in the Hamiltonian $\hat{H}^{(0)}$:
\begin{equation}
 \hat{H}^{(0)} \big| \psi_0^{(0)} \big> = \varepsilon_0 \big| \psi_0^{(0)} \big>
 \label{EQ:PSI_VAR_0}
\end{equation}
Expanding the wavefunction $\left| \psi_0 \right>$ for the target state into a perturbation series where $\hat{V}$ is the 
perturbation and retaining only non-zero terms yields
\begin{eqnarray}
 \nonumber
 \left| \psi_0 \right> &=& \left| \psi_0^{(0)} \right> + \sum\limits_{k\ne 0}\, \frac{V_{k0}}{\varepsilon_0 -\varepsilon_k} 
                           \left| \psi_k^{(0)} \right>
        -\frac{1}{2} \sum\limits_{k \ne 0}\, \frac{ |V_{k0}|^2 }{(\varepsilon_0 -\varepsilon_k)^2} \left| \psi_0^{(0)} \right> \\
    &&  + \sum\limits_{k,l \ne 0}\, \frac{ V_{kl} V_{l0} }{(\varepsilon_0 -\varepsilon_k)(\varepsilon_0 -\varepsilon_l)}
                   \left| \psi_k^{(0)} \right> + {\cal{O}}(\hat{V}^3)
 \label{EQ:PT}
\end{eqnarray}
where $V_{kl} = \left< \psi_k^{(0)} | \hat{V} | \psi_l^{(0)} \right>$ and $\varepsilon_m$ is an unperturbed energy. 
If terms ${\cal{O}}(\hat{V}^3)$ are omitted from Eq. (\ref{EQ:PT}) and the resulting truncated expansion is used to evaluate
the expectation value $\left< \psi_0 \right| \hat{H}_{\text{HF}} \left| \psi_0 \right>$ then the third-order expressions
from Refs. \cite{Angstmann_Dzuba_Flambaum_PRA2006} and \cite{schwartz_SPT_1959} arise. In the present work, however,
${\cal{O}}(\hat{V}^3)$ is implicitly included in Eq. (\ref{EQ:PSI_VAR}) which leads to the presence of all orders of
the external electric field in Eq. (\ref{EQ:A_CONST}) and, therefore, also in Eq. (\ref{EQ:K_DEF}) defining the hyperfine
Stark coefficient.

\subsection{Scalar and Tensor Static Polarizabilites}

An electric dipole polarizability for a state labelled $J,M_J$ can be given 
\cite{angel_sandars_pol1,Manakov_Ovsyannikov_1979,Chen_Raithel_2015} 
in terms of scalar ($\alpha_0$) and tensor ($\alpha_2$) static polarizabilites as follows:
\begin{equation}
 \alpha_{J,M_J} = \alpha_0(J) + \alpha_2(J) \frac{3M_J^2 - J(J+1)}{J(2J-1)}
 \label{EQ:ALPHA_SCTEGEN}
\end{equation}
where $\alpha_0$ and $\alpha_2$ are functions of $J$ only.
If $\alpha_{J,M_J}$ and $\alpha_{J,M_J'}$ with $M_J \ne M_J'$ are known then the resulting system of Eqs. 
(\ref{EQ:ALPHA_SCTEGEN})
can be inverted and $\alpha_0$ and $\alpha_2$ become calculable functions of $\alpha_{J,M_J}$ and $\alpha_{J,M_J'}$.
This fact will be exploited in order to compare with polarizability results given in terms of scalar and tensor
polarizabilities in cases where $J > \frac{1}{2}$.

This is the case for the Thulium atom ground levels. In the calculations presented below $\alpha_{J,M_J}$ are
calculated using the finite-field method (see subsection \ref{SUBSEC:FF}). $\alpha_0(J)$ and $\alpha_2(J)$ are
then obtained by inverting Eqs. (\ref{EQ:ALPHA_SCTEGEN}). Finally, the differential static scalar polarizability 
in this case results from
\begin{equation}
 \Delta\alpha_0 = \alpha_0(J_u) - \alpha_0(J_l)
 \label{EQ:POL_DIFF}
\end{equation}
and the hyperfine Stark shift parameter can be obtained from this as
\begin{equation}
 k = -\frac{1}{2}\, \Delta\alpha_0
\end{equation}

\section{Applications and results}
\label{SEC:RESULTS}
\subsection{Technical Details}

\subsubsection{External Electric Field}

The external electric field $E_z$ has to be chosen small enough to assure that $k$ as defined in Eq. (\ref{EQ:K_DEF})
does not vary with $E_z$, but also large enough to ensure that the field-dependent change in $\psi(E_z)$ is sufficiently
greater than the convergence threshold chosen for the wavefunction. Typically, this is achieved for $E_z = 10^{-3}$ \au\ 
which is the value used in the present direct calculations. In addition, a region around $E_z = 10^{-3}$ \au\ has been 
explored which allows for an estimate of the uncertainty in $k$ due to deviations from quadratic dependency of the
field-dependent total energy.

\subsubsection{Atomic Basis Sets}

For all three atoms Gaussian basis sets are used. The Rb atom is described by the uncontracted Dyall vTZ set with 
$28s,20p,12d,1f$ primitive functions and the aug-cc-pwCVQZ-X2C set from
the EMSL library \cite{EMSL-basis2019} that has been singly densified in the $s$ and $p$ spaces and augmented by one dense 
and one diffuse function as described in Ref. \cite{Hubert_Fleig_2022}. The full uncontracted QZ+ set then comprises
$77s,59p,21d,5f,3g$ functions. The Cs atom is described by the uncontracted Dyall vTZ set with $31s,24p,15d,1f$ primitive
functions and the vQZ basis \cite{dyall_s-basis} with 
$5s5p6s6p$ correlating exponents added. Like for Rb the $s$ and $p$ spaces have been densified once and augmented. The 
final uncontracted QZ+ set comprises $75s,61p,19d,3f,1g$ functions.

For the thulium atom two different basis sets \cite{dyall_4f} are employed: 1) The uncontracted Dyall vTZ set 
including all $[4f/6s/5s,p,d]$ correlating and $4f$ dipole-polarizing functions adding up to $[30s,24p,18d,13f,4g,2h]$ 
functions (in the following denoted as TZ). 2) The uncontracted Dyall cvQZ basis including all $[4f/6s/5d]$ correlating 
and $4f$ dipole-polarizing functions comprised by a total of $[35s,30p,19d,16f,6g,4h,2i]$ functions (QZ).

The densification and augmentation in the $s$ and $p$ spaces in some cases serves to test a highly accurate basis set for 
describing the $s$-$p$ mixing which is predominantly responsible for the energy shifts when an external electric 
field is applied to hyperfine states.

\subsubsection{Atomic Wavefunctions}

All atoms considered here have an odd number of electrons. In addition, the interaction with $E_z$ breaks the full 
rotational symmetry of the atom which means that only $M_J$ remains as an exact electronic quantum number (for
$\psi(E_{\text{ext}})$ in Eq. (\ref{EQ:PSI_VAR})).
Atomic wavefunctions are calculated in the $\left|M_J\right| = \Omega$ irreducible representations of the 
$C_{\infty v}^*$
double point group. The target atom is placed at the origin of the reference frame and a ghost atom with neither
electronic basis functions nor nuclear charge is placed at a finite distance along the axis of the external $E$ 
field in order to allow for the inclusion of the external-field Hamiltonian in linear-symmetry calculations.
The ghost atom introduces no physical interaction.

For solving Eq. (\ref{EQ:PSI_VAR}) the KRCI module \cite{knecht_luciparII} of the DIRAC program package 
\cite{DIRAC_JCP} is used. In a first step the Dirac-Coulomb-Hartree-Fock equations are solved where the Hamiltonian in 
Eq. (\ref{EQ:HAMILTONIAN}) is employed. This model will be abbreviated as DCHF. The atomic spinors are optimized by 
diagonalizing a Fock operator where a fractional occupation of $f = \frac{m}{n}$ per spinor in the defined valence
shells is used. Here $m$ is the number of electrons and $n$ is the number of spinors the respective shell comprises.
These spinors are thus obtained for the electric potential of the neutral atom.

Acronyms are used for brevity in defining atomic correlated wavefunctions from the second step.
As an example, SDT9\_10au stands for Single, Double, Triple replacements relative to the DCHF reference state
where 9 electrons occupying the outermost shells in the DCHF reference state are taken into account in the
correlation expansion and the complementary space of virtual spinors is truncated at 10 \au  An acronym SD8\_SDT9
means that up to two holes are allowed in the shells occupied by 8 electrons and up to triple excitations into
the virtual spinors are allowed from the combined shells occupied by the 9 electrons in the reference state. 
In the case of an alkali atom this means that up to double excitations from the $(n-1)s~ (n-1)p$ shells and 
up to triple excitations from the combined $(n-1)s~ (n-1)p$ and $ns$ shells are included in the wavefunction
expansion. Thus, the model SD8\_SDT9 comprises a subset of the determinants of the model SDT9, where in the 
latter {\it{all}} triple excitations from the $(n-1)s~ (n-1)p$ and $ns$ shells are included.

\subsubsection{Finite-Field (FF) Method}
\label{SUBSEC:FF}

Static dipole polarizabilities $\alpha_J$ are calculated by fitting finite-field electronic energies for four field 
points with $E \in \{ 0, 0.00025, 0.0005, 0.001\}$ a.u. to a polynomial and extracting the second derivative of the fitted
function at zero field which is proportional to the static dipole polarizability, see also Refs. \cite{fleig_halo,fleig_pol13}.


\subsection{$^{87}$Rb}

\subsubsection{Static Electric Dipole Polarizability $\alpha_D$}

As a corroboration of the present method of calculating polarizabilites a comparison with literature results for scalar and
tensor polarizabilities is drawn. Using Eq. (\ref{EQ:ALPHA_SCTEGEN}) the $M_J$-dependent values of $\alpha_D$ are obtained 
from the equations
\begin{eqnarray}
 \nonumber
 \alpha_D\left(^2P_{3/2,1/2}\right) &=& \alpha_0\left(^2P_{3/2}\right) - \alpha_2\left(^2P_{3/2}\right) \\
 \alpha_D\left(^2P_{3/2,3/2}\right) &=& \alpha_0\left(^2P_{3/2}\right) + \alpha_2\left(^2P_{3/2}\right)
 \label{EQ:ALPHA_D_MJ}
\end{eqnarray}
where $\alpha_0$ is the scalar polarizability and $\alpha_2$ is the tensor polarizability. This allows for calculating the
$M_J$-dependent polarizabilities from the results in Ref. \cite{Arora_Sahoo_2012} and these are given in Table \ref{TAB:ALPHA}.
The inverted equations (\ref{EQ:ALPHA_D_MJ}) read
\begin{eqnarray}
 \nonumber
 \alpha_0\left(^2P_{3/2}\right) &=& \frac{1}{2} \left[ \alpha_D\left(^2P_{3/2,3/2}\right) + \alpha_D\left(^2P_{3/2,1/2}\right) \right] \\
 \alpha_2\left(^2P_{3/2}\right) &=& \frac{1}{2} \left[ \alpha_D\left(^2P_{3/2,3/2}\right) - \alpha_D\left(^2P_{3/2,1/2}\right) \right]
 \label{EQ:ALPHA_D_SCATEN}
\end{eqnarray}
This, in turn, allows for calculating the scalar and tensor polarizabilities from the present $M_J$-dependent values, the results
of which are also given in Table \ref{TAB:ALPHA}.

\begin{table}[h]

 \begin{center}
       \caption{\label{TAB:ALPHA} 
         $M_J$-dependent static electric dipole polarizabilities $\alpha_D$ [a.u.] calculated through the FF method 
         for states $^ML_{J,M_J}$ where $M = 2S +1$ is the spin multiplicity and scalar ($\alpha_0$) and tensor
         ($\alpha_2$) polarizabilities for the $^2P_{3/2}$ state; for the electronic ground state $^2S_{1/2,1/2}$,
         $\alpha_D\left(^2S_{1/2,1/2}\right) = \alpha_0\left(^2S_{1/2}\right)$.
       }

 \vspace*{0.3cm}
 \begin{tabular}{l|c|ccc|cc}
                       & $5s^1$          & \multicolumn{5}{c}{$5p^1$}                              \\
Model& $^2S_{1/2,1/2}$ & $^2P_{1/2,1/2}$ & $^2P_{3/2,1/2}$ & $^2P_{3/2,3/2}$ & $\alpha_0(^2P_{3/2})$ & $\alpha_2(^2P_{3/2})$ \\ \hline
  DCHF                        &   $485.3$       &                 &                 &              &           &            \\
  QZ+/SDT9/10au               &   $333.0$       &   $820.3$       &   $1047.9$      &   $719.3$    &  $883.6$  &  $-164.3$  \\ \hline
  Experiment (cited in Ref. \cite{Arora_Sahoo_2012}) & $318.79$ & $810.6$ & $1020$  &   $694$      &  $857.0$  &  $-163  $  \\
  Other theory (\cite{Arora_Sahoo_2012}) & $318.3$ & $810.5$      &   $1033.9$      &   $702.1$    &  $868.0$  &  $-165.9$ 
 \end{tabular}
 \end{center}
\end{table}

For the $^2P_{3/2}$ level the present results for $\alpha_0$ and $\alpha_2$ which are derived from $M_J$-dependent calculations
as described above differ from the experimental values by only $1-3$\%. Further improvements could be made by modifying the
employed electronic-structure model, but for the present case the obtained correspondence serves as a sufficient proof of
principle for the applied method.

\subsubsection{Hyperfine Stark Effect}

For the electronic ground state corresponding to the valence configuration $4s^1$ the total electronic angular momentum
quantum number is ${J=1/2}$. The considered isotope has $I = 3/2$ and the resulting hyperfine quantum levels are 
denoted as $F_u = 2$ and $F_l = 1$. The fractional occupation in the DCHF calculation is $f=1/2$.

Results for the hyperfine Stark coefficient are shown in Table \ref{TAB:A_RB} for various electronic-structure
models and are compared with experimental and theoretical literature results. 
As a general effect, the hyperfine constant of a given atom in an electronic $S$ state diminishes when an external $E$
field is included. This observation is explained by the fact that the $E$ field partially shifts spin density
from $s$ wave to $p$ wave character in the atomic ground state (through $s$-$p$ mixing in the polarized atom), thus 
reducing the hyperfine interaction.

The hyperfine Stark coefficient in mean-field approximation (DCHF, both basis sets) differs from the experimental result 
by more than $20$\% and lowest-order electron correlation effects from the $4s,4p,5s$ Rb shells even increase this discrepancy 
(model SD9, both basis sets).  
Upon including combined triple excitations from the $4s,4p$ shells and the $5s$ valence shell (model TZ/SD8\_SDT9) a 
strong correction to $k$ is obtained. 
The inclusion of full triple excitations (model TZ/SDT9), \ie\ including those from the $4s,4p$ shells, yields another
large change in $k$.
Close agreement with the result obtained by Safronova et al. \cite{Safronova_BBR_2009} is achieved at this level
of calculation. This is not surprising since the wavefunction model used in Ref. \cite{Safronova_BBR_2009} for obtaining
the cited result is a linearized coupled cluster expansion including up to (perturbative) valence triple excitations 
which is similar to the present SDT9 model.

\begin{table}[h]
   \caption{\label{TAB:A_RB} Stark hyperfine coefficient $k$ for $^{87}$Rb with nuclear magnetic moment 
            $\mu = 2.75131 \mu_N$ \cite{ISOLDE_BWeffect_1993}
           }

 \begin{tabular}{l|c}
	 Model                                                     & $k$ [$10^{-10}$ Hz/((V/m)$^2$)]   \\ \hline
         TZ/DCHF                                                   &   $-1.522$                        \\
         TZ/S8\_SD9/10au                                           &   $-1.138$                        \\ 
         TZ/SD9/10au                                               &   $-1.618$                        \\ 
         TZ/SD8\_SDT9/10au                                         &   $-1.220$                        \\ 
         TZ/SDT9/10au                                              &   $-1.273$                        \\ 
         TZ/SDT8\_SDTQ9/10au                                       &   $-1.159$                        \\ 
         TZ/SDTQ9/10au                                             &   $-1.228$                        \\ 
         TZ/SDTQ8\_SDTQQ9/10au                                     &   $-1.158$                        \\
         TZ/SDTQQ9/10au                                            &   $-1.161$                        \\ 
         TZ/SD18\_SDT19/10au                                       &   $-1.253$                        \\ 
         TZ/SD26\_SDT27/10au                                       &   $-1.295$                        \\
         TZ/SD34\_SDT35/10au                                       &   $-1.299$                        \\ 
         TZ/SD36\_SDT37/10au                                       &   $-1.301$                        \\ \hline  
         QZ+/DCHF                                                  &   $-1.546$                        \\
         QZ+/SD9/10au                                              &   $-1.662$                        \\ 
         QZ+/SD9/30au                                              &   $-1.669$                        \\
         QZ+/SD8\_SDT9/30au                                        &   $-1.202$                        \\
         QZ+/SDT9/10au                                             &   $-1.265$                        \\
         QZ+/SDT9/30au                                             &   $-1.271$                        \\ \hline
         {\bf{Final}}                                              &   ${\boldsymbol{-1.234 \pm 0.0223}}$  \\ \hline
         Exp. \cite{Mowat_Li_1972}                                 &   $-1.23(3)$                      \\
         Safronova et al. \cite{Safronova_BBR_2009}                &   $-1.272$\footnotemark[1]        \\
         Angstmann et al. \cite{Angstmann_Dzuba_Flambaum_PRA2006}  &   $-1.24(1)$
 \end{tabular}
 \footnotetext[1]{preliminary value}
\end{table}

\begin{figure}
 \caption{Stark hyperfine coefficient $k$ for the electronic ground state of the rubidium atom using various
          electronic-structure models (black: present QZ+, blue: present TZ) and compared with other theoretical results
          and experiment, including the experimental uncertainty; 
          for three of the models the evolution of the result with the number of CI iterations is also shown.}
 \label{FIG:RB}
 \begin{center}
  \includegraphics[angle=0,width=18.0cm]{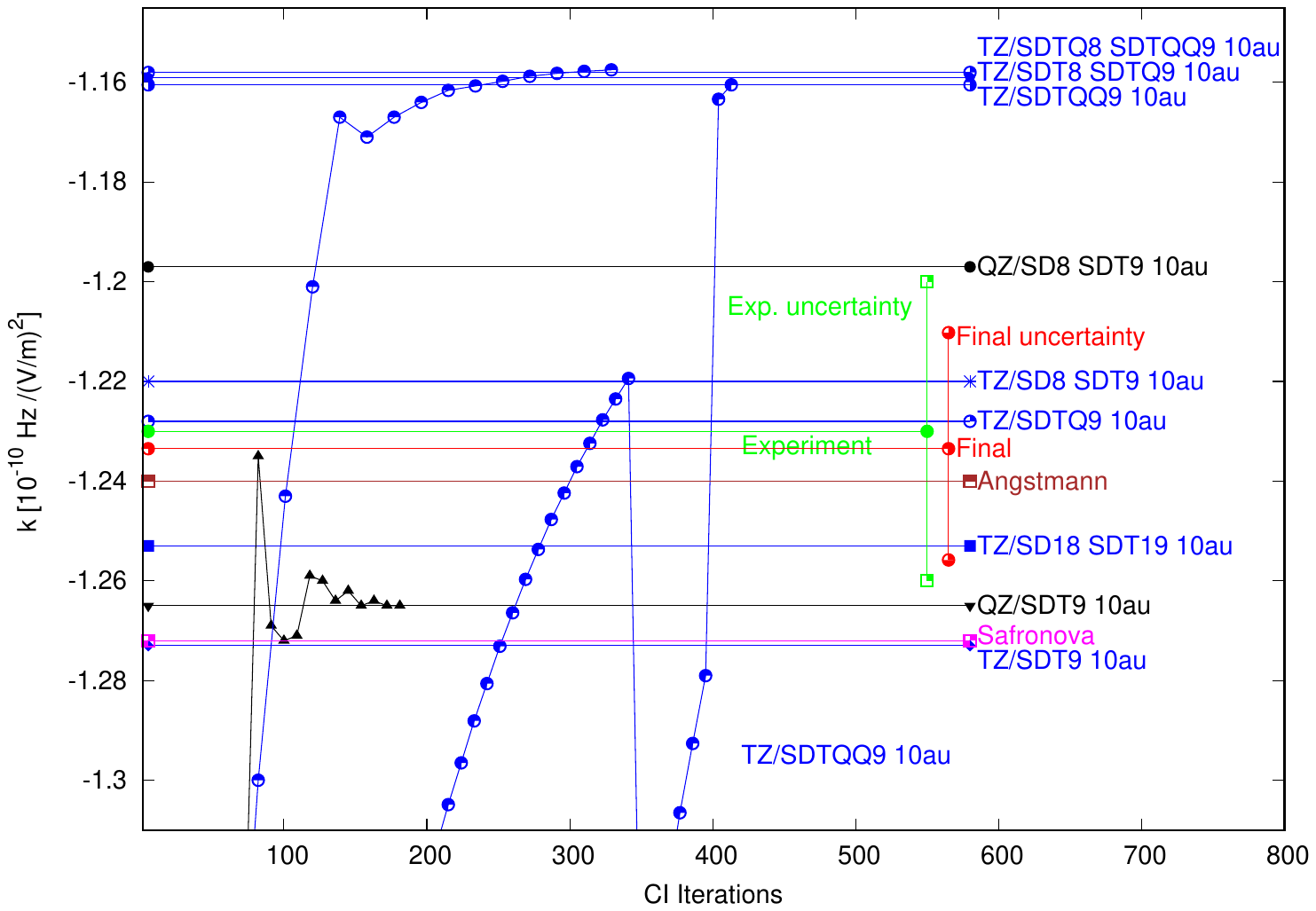}
 \end{center}
\end{figure}

It becomes clear that higher excitation ranks in the wavefunction expansion have a much greater impact on the hyperfine
Stark coefficient than improvements in the atomic basis set. It is thus attempted to systematically converge $k$ with respect 
to the former effects. 
The evolution of the results for these systematically improved models is displayed graphically in Fig. \ref{FIG:RB} for ease
of comparison.

%
%

Along the improving model series SD9 -- SD8\_SDT9 -- SDT9 -- SDT8\_SDTQ9 -- SDTQ9 $\ldots$ results strongly oscillate, even after
having included combined quadruple excitations. However, the partial series of corresponding models 
SD8\_SDT9 -- SDT8\_SDTQ9 -- SDTQ8\_SDTQQ9 does lead to a sufficiently converged result. Likewise, the partial series
SD9 -- SDT9 -- SDTQ9 -- SDTQQ9 also reaches a sufficiently converged value that is not far from the result with the model
SDTQ8\_SDTQQ9. It can be inferred that the result from the model SDTQQ9 that includes full quintuple excitations for the $9$
outermost electrons is near the Full CI result in the TZ basis set.   

The final value is obtained as follows. The result from the model SDTQQ9 serves as the base value and corrections from correlations
of inner-shell electrons and from the atomic basis set are added. In detail,
\begin{eqnarray}
 \nonumber
 k({\text{final}}) &=& k({\text{TZ/SDTQQ9/10au}})  \\
 \nonumber
                &&  + k({\text{TZ/SD36\_SDT37/10au}}) - k({\text{TZ/SD8\_SDT9/10au}})  \\
                &&  + k({\text{QZ+/SDT9/10au}}) - k({\text{TZ/SDT9/10au}})
\end{eqnarray}
This result differs from the experimental central value by only about $0.3$\% and is the theoretical result that comes closest
to the experimental value. The uncertainty for the present final value is obtained by addition of individual uncertainties for
interelectron correlation effects, the basis-set approximation and the approximation due to the use of the Dirac-Coulomb
Hamiltonian operator (which neglects the Breit interaction and QED effects) for obtaining the atomic wavefunctions. The former 
two uncertainties are obtained from the difference
between the most elaborate and the second most elaborate models, respectively. The latter uncertainty is estimated to be
1\%.

\subsection{$^{133}$Cs}

This important isotope of cesium has received considerable attention in the past.
For the electronic ground state corresponding to the valence configuration $6s^1$ the total electronic angular momentum
quantum number is ${J=1/2}$. The considered isotope has $I = 7/2$ and the resulting hyperfine quantum levels are
denoted as $F_u = 4$ and $F_l = 3$. The fractional occupation in the DCHF calculation is $f=1/2$. 

Results for $^{133}$Cs are compiled in Table \ref{TAB:A_CS}.
As in the case of $^{87}$Rb the mean-field result for
$k$ is too large on the absolute, compared with the cited reference values, by around $40$\%. Again, the inclusion of
lowest-order electron correlation effects from the outermost atomic shells even increases this deviation. 
It is here in addition and for the model QZ+/SD9 shown that the consideration of virtual spinors of very high energy 
(up to $1000$ \au) in the wave-function expansion
does not affect the results significantly. Also as for $^{87}$Rb combined triple excitations from the outermost shells
yield a very important correction to $k$. Including full triple excitations (model QZ+/SDT9/30au) yields a result that deviates
from the most accurate literature results by roughly $+9$\%. Accounting for combined quadruple excitations quenches
the deviation to $-7$\% where again, as in the Rb atom, the correction slightly overshoots the exact
value. This latter model comprises about $1.9$ billion ($10^9$) terms in the CI expansion when the QZ+ basis set is
used.

\begin{table}[h]
	\caption{\label{TAB:A_CS} Stark hyperfine frequency shift $k$
                 for $^{133}$Cs ($5s_{1/2}$), $I = 3.5$, $E_z = 0.001$ a.u., hyperfine level quantum numbers are 
                 $F_u = 4$, $F_l = 3$.
	}

 \vspace*{0.3cm}
 \begin{tabular}{l|l}
    Model                                                  & $k$ [$10^{-10}$ Hz/((V/m)$^2$)] \\ \hline
   TZ/DCHF                                                 &   $-3.164$     \\
   TZ/SD9/10au                                             &   $-3.272$     \\
   TZ/SD8\_SDT9/10au                                       &   $-2.319$     \\
   TZ/SDT9/10au                                            &   $-2.433$     \\
   TZ/SDT8\_SDTQ9/10au                                     &   $-2.153$     \\
   TZ/SDTQ9/10au                                           &   $-2.321$     \\
   TZ/SDTQ8\_SDTQQ9/10au                                   &   $-2.156$     \\
   TZ/SDTQQ9/10au                                          &   $-2.161$     \\
   TZ/SD19/10au                                            &   $-3.335$     \\
   TZ/SDT19/10au                                           &   $-2.516$     \\
   TZ/SDT27/10au                                           &   $-2.595$     \\
   TZ/SD27/10au                                            &   $-3.395$     \\
   TZ/SD37/10au                                            &   $-3.396$     \\           
   TZ/SD45/10au                                            &   $-3.380$     \\ \hline
   QZ+/DCHF                                                &   $-3.220$     \\
   QZ+/SD9/10au                                            &   $-3.375$     \\
   QZ+/SD9/30au                                            &   $-3.379$     \\ 
   QZ+/SD9/1000au                                          &   $-3.384$     \\
   QZ+/SD8\_SDT9/10au                                      &   $-2.308$     \\
   QZ+/SD8\_SDT9/30au                                      &   $-2.311$     \\
   QZ+/SDT9/10au                                           &   $-2.468$     \\
   QZ+/SDT9/30au                                           &   $-2.469$     \\
   QZ+/SDT8\_SDTQ9/10au                                    &   $-2.112$     \\ \hline
   {\bf{Final}}                                            &   ${\boldsymbol{-2.347 \pm 0.084}}$    \\ \hline
   Exp.\cite{Simon_PRA1998}                                &   $-2.271(4)$  \\
   Exp.\cite{Godone_PRA2005}                               &   $-2.05(5)$   \\
   Safronova et al.\cite{beloy_derevianko_PRL2006}         &   $-2.271(8)$  \\
   Angstmann et al.\cite{Angstmann_Dzuba_Flambaum_PRA2006} &   $-2.26(2)$  
 \end{tabular}

\end{table}

Even higher excitation ranks can with the present method only be treated in the smaller TZ basis set. There, even a
calculation with full quadruple excitations does not yet deliver a converged result (model TZ/SDTQ9). A converged
result, however, is obtained when at least combined quintuple excitations are included in the wavefunction expansion.

\begin{figure}
 \caption{Stark hyperfine coefficient $k$ for the electronic ground state of the cesium atom using various
          electronic-structure models (black: present QZ, blue: present TZ) and compared with other theoretical results
          and experiment, including the experimental uncertainty; 
          for two of the models the evolution of the result with the number of CI iterations is also shown.}
 \label{FIG:CS}
 \begin{center}
  \includegraphics[angle=0,width=18.0cm]{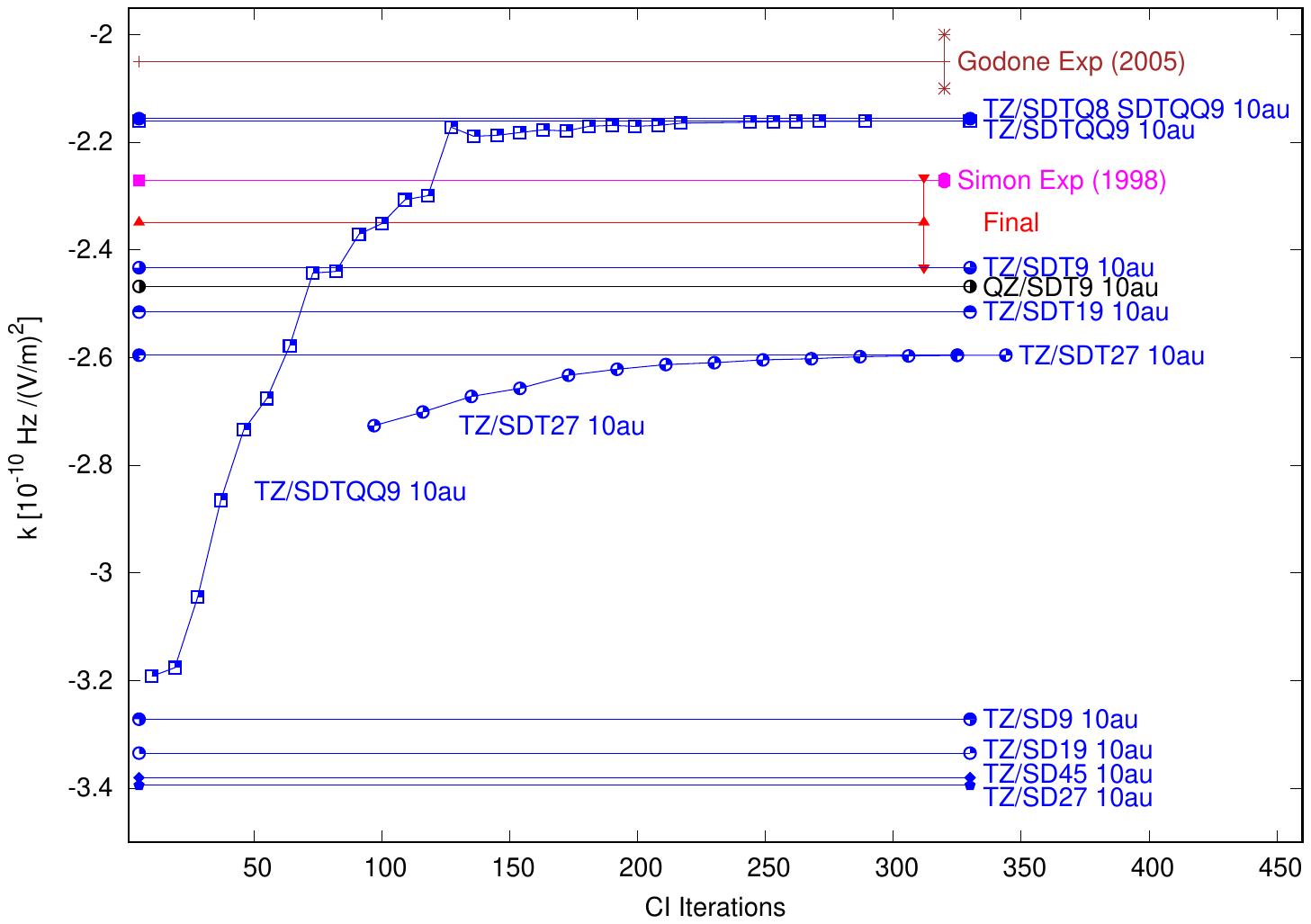}
 \end{center}
\end{figure}

Fig. \ref{FIG:CS} displays a selection of results supporting the present discussion. Correlations among and with the 
electrons from inner shells down to and including the $3s$ shell (model SD45) lead to a small increase of $k$, on the
absolute. A very similar correction is obtained when full triple excitations are included in these expansions (models
SDT9 and SDT27). This suggests that there is a negligible effect from including even higher excitation ranks
in order to obtain corrections from inner-shell electron correlations.

%

The final present value of the hyperfine Stark shift for Cs is, therefore, obtained as follows. The base value is taken
from the model TZ/SDTQ8\_SDTQQ9. This result is corrected by inner-shell correlations from the $4s,4p,4d$ shells at the SDT 
level and for inner-shell correlations from the $3s,3p,3d$ shells at the SD level. A small basis-set correction is added
from the SDT9 model as well as a correction from including virtual spinors up to 1000 a.u. (model QZ+/SD9). Mathematically, 
the corresponding evaluations are
\begin{eqnarray}
 \nonumber
 k({\text{final}}) &=& k({\text{TZ/SDTQ8\_SDTQQ9/10au}})  \\
 \nonumber
                &&  + k({\text{TZ/SDT27/10au}}) - k({\text{TZ/SDT9/10au}})  \\
 \nonumber
                &&  + k({\text{TZ/SD45/10au}}) - k({\text{TZ/SD27/10au}})  \\
 \nonumber
                &&  + k({\text{QZ+/SD9/1000au}}) - k({\text{QZ+/SD9/10au}})  \\
                &&  + k({\text{QZ+/SDT9/10au}}) - k({\text{TZ/SDT9/10au}})
\end{eqnarray}
The uncertainty on this final result is estimated in the same way as has been done for the above Rb atom. 

The present final result is compatible with the measurement by Simon {\it{et al.}} from 1998 \cite{Simon_PRA1998} and 
incompatible with the measurement by Godone {\it{et al.}} from 2005 \cite{Godone_PRA2005}. The present result is also compatible 
with the theoretical results by Safronova {\it{et al.}} \cite{beloy_derevianko_PRL2006} and by Angstmann {\it{et al.}} 
\cite{Angstmann_Dzuba_Flambaum_PRA2006}. The result by Godone {\it{et al.}} is in conflict with all other experimental
and theoretical results. The present result supports previous theoretical values as well as the measurement by Simon 
{\it{et al.}}

\subsection{$^{169}$Tm}

With confidence in the method established in the aforegoing sections it is now applied to an atom where high-level 
theoretical reference results are not available and which has a more complex electronic structure. 
In this case the fractional occupation in the DCHF calculation is $f=15/16$ where $m=15$ represents the thirteen $4f$
electrons plus the two $6s$ electrons. This averaging was required to assure proper convergence of the DCHF
wavefunction.

\subsubsection{$^2F_{7/2}(F=4)$--$^2F_{5/2}(F=3)$ clock transition}

The clock transition discussed in 
Refs. \cite{NatComm_Tm_2019,PRA_Tm_2016,PRA_Tm_2020} comprises a hyperfine component ($F_l = 4$) of the ground electronic 
state $^2F_{7/2}$ and a hyperfine component ($F_u = 3$) of the first electronically excited state $^2F_{5/2}$. 
In this case the Stark coefficient is dominated by the polarizability difference between the two respective electronic
states and Eq. (\ref{EQ:POL_DIFF}) applies.

Individual static polarizabilities $\alpha_J$ are calculated by using the FF method and are given in Table \ref{TAB:ALPHA_TM7}.
Using Eq. (\ref{EQ:ALPHA_SCTEGEN}) the $M_J$-dependent values of $\alpha_D$ are in the case of thulium states with $J = 7/2$ obtained 
from the inverted equations
\begin{eqnarray}
 \alpha_0\left(^2F_{7/2}\right) &=& \frac{7}{6} \alpha_D\left(^2F_{7/2,5/2}\right) -\frac{1}{6} \alpha_D\left(^2F_{7/2,7/2}\right)  \\
 \alpha_2\left(^2F_{7/2}\right) &=& \frac{7}{6} \left[ \alpha_D\left(^2F_{7/2,7/2}\right) - \alpha_D\left(^2F_{7/2,5/2}\right) \right]  \\
 \alpha_0\left(^2F_{7/2}\right) &=& \frac{5}{2} \alpha_D\left(^2F_{7/2,3/2}\right) -\frac{3}{2} \alpha_D\left(^2F_{7/2,1/2}\right)  \\
 \alpha_2\left(^2F_{7/2}\right) &=& \frac{7}{2} \left[ \alpha_D\left(^2F_{7/2,3/2}\right) - \alpha_D\left(^2F_{7/2,1/2}\right) \right]
 \label{EQ:ALPHA_D_SCATEN_TM7}
\end{eqnarray}
results for which are also given in Table \ref{TAB:ALPHA_TM7}.
%

\begin{table}[h]

 \begin{center}
       \caption{\label{TAB:ALPHA_TM7}
         $M_J$-dependent static electric dipole polarizabilities $\alpha_D$ [a.u.] calculated through the FF method 
         for states $^ML_{J,M_J}$ where $M = 2S +1$ is the spin multiplicity and scalar ($\alpha_0$) and tensor
         ($\alpha_2$) polarizabilities for the $^2F_{7/2}$ multiplet
       }

 \vspace*{0.3cm}
 \begin{tabular}{l|cccc|cc|c}
     &  \multicolumn{4}{c|}{$\alpha_D$} &  &  &   \\ 
Model& $^2F_{7/2,1/2}$ & $^2F_{7/2,3/2}$ & $^2F_{7/2,5/2}$ & $^2F_{7/2,7/2}$ & $\alpha_0(^2F_{7/2})$ & $\alpha_2(^2F_{7/2})$ &
     calc. from $\{M_J\}$ \\ \hline
  TZ/DCHF                 &   $186.4554$ &   $185.7002$ &  $        $   &              &   $184.5674$ &  $-2.643$  & $1/2,3/2$    \\ 
  TZ/SD15/6au             &   $163.9967$ &   $163.2578$ &  $        $   &              &   $162.1495$ &  $-2.586$  & $1/2,3/2$    \\ 
  TZ/SD13\_SD15/6au       &   $163.9897$ &   $163.2515$ &  $        $   &              &   $162.1442$ &  $-2.584$  & $1/2,3/2$    \\ 
  TZ/SD13\_SDT15/6au      &   $161.9754$ &   $161.3215$ &  $        $   &              &   $160.3407$ &  $-2.289$  & $1/2,3/2$    \\ 
  TZ/SD13\_SDTsppdQ15/6au &   $167.2591$ &   $166.5212$ &  $        $   &              &   $165.4144$ &  $-2.583$  & $1/2,3/2$    \\ 
  QZ/DCHF                 &   $186.6210$ &   $185.8505$ &  $        $   &              &   $184.6948$ &  $-2.697$  & $1/2,3/2$    \\ 
  QZ/SD15/10au            &   $161.2430$ &   $160.4891$ &  $158.9842$   &  $156.7269$  &   $159.3583$ &  $-2.639$  & $1/2,3/2$    \\ 
                          &              &              &  $        $   &  $        $  &   $159.3604$ &  $-2.634$  & $5/2,7/2$    \\ 
  QZ/SDT15/10au           &   $160.8884$ &   $160.2111$ &               &              &   $159.1952$ &  $-2.371$  & $1/2,3/2$    \\ 
  QZ/SD23/10au            &   $151.2078$ &   $150.4219$ &  $        $   &              &   $149.2431$ &  $-2.751$  & $1/2,3/2$    \\ 
  QZ/SD33/10au            &   $151.8781$ &   $151.0859$ &  $149.5087$   &  $147.1326$  &   $149.8976$ &  $-2.773$  & $1/2,3/2$    \\ 
                          &              &              &               &              &   $149.9047$ &  $-2.772$  & $5/2,7/2$    \\ \hline
  Experiment              &              &              &               &              &   $130 \pm 16$ \cite{Ma_Tl_pol} & &      \\
  Recommended             &              &              &               &              &   $144 \pm 15$ \cite{schwerdt_nagle_pol_atoms_2018} & & 
 \end{tabular}
 \end{center}
\end{table}
%
Both the scalar and tensor polarizabilites are numerically invariant to the choice of $M_J$ components used for their 
calculation up to two digits after the decimal point. This is shown for one case using the model QZ/SD33/10au. All 
scalar and tensor polarizabilites are therefore calculated from $\alpha_{7/2,3/2}$ and $\alpha_{7/2,1/2}$.

For $J = 5/2$ states of thulium the inverted equations read
\begin{eqnarray}
 \alpha_0\left(^2F_{5/2}\right) &=& \frac{4}{3} \alpha_D\left(^2F_{5/2,3/2}\right) -\frac{1}{3} \alpha_D\left(^2F_{5/2,1/2}\right)  \\
 \alpha_2\left(^2F_{5/2}\right) &=& \frac{5}{3} \left[ \alpha_D\left(^2F_{5/2,3/2}\right) - \alpha_D\left(^2F_{5/2,1/2}\right) \right]  \\
 \alpha_0\left(^2F_{5/2}\right) &=& \frac{1}{6} \alpha_D\left(^2F_{5/2,5/2}\right) +\frac{5}{6} \alpha_D\left(^2F_{5/2,3/2}\right)  \\
 \alpha_2\left(^2F_{5/2}\right) &=& \frac{5}{6} \left[ \alpha_D\left(^2F_{5/2,5/2}\right) - \alpha_D\left(^2F_{5/2,3/2}\right) \right] 
 \label{EQ:ALPHA_D_SCATEN_TM5}
\end{eqnarray}
This, in turn, allows for calculating the scalar and tensor polarizabilities from the present $M_J$-dependent values, the results
of which are also given in Table \ref{TAB:ALPHA_TM5}.

\begin{table}[h]
 \begin{center}
       \caption{\label{TAB:ALPHA_TM5} 
         $M_J$-dependent static electric dipole polarizabilities $\alpha_D$ [a.u.] calculated through the FF method 
         for states $^ML_{J,M_J}$ where $M = 2S +1$ is the spin multiplicity and scalar ($\alpha_0$) and tensor
         ($\alpha_2$) polarizabilities for the $^2F_{5/2}$ multiplet
       }
 \vspace*{0.3cm}
 \begin{tabular}{l|ccc|cc|c}
     &  \multicolumn{3}{c|}{$\alpha_D$} &  &  &   \\ 
Model& $^2F_{5/2,1/2}$ & $^2F_{5/2,3/2}$ & $^2F_{5/2,5/2}$ & $\alpha_0(^2F_{5/2})$ & $\alpha_2(^2F_{5/2})$ & calc. from $\{M_J\}$  \\ \hline
  TZ/DCHF                 &   $186.3443$    &   $185.0430$    &   $       $  &  $184.6092$   &  $-2.169$   & $1/2,3/2$    \\ 
  TZ/SD15/6au             &   $163.9999$    &   $162.7315$    &   $       $  &  $162.3087$   &  $-2.114$   & $1/2,3/2$    \\ 
  TZ/SD13\_SD15/6au       &   $163.9924$    &   $162.7252$    &   $       $  &  $162.3028$   &  $-2.112$   & $1/2,3/2$    \\ 
  TZ/SD13\_SDT15/6au      &   $161.9244$    &   $160.8021$    &   $       $  &  $160.4280$   &  $-1.871$   & $1/2,3/2$    \\ 
  TZ/SD13\_SDTsppdQ15/6au &   $167.1381$    &   $165.8626$    &   $       $  &  $165.4374$   &  $-2.126$   & $1/2,3/2$    \\ 
  QZ/DCHF                 &   $186.5082$    &   $185.1807$    &   $       $  &  $184.7382$   &  $-2.213$   & $1/2,3/2$    \\ 
  QZ/SD15/10au            &   $161.2017$    &   $159.9065$    &   $       $  &  $159.4748$   &  $-2.159$   & $1/2,3/2$    \\ 
  QZ/SDT15/10au           &   $160.7452$    &   $159.5866$    &              &  $159.2004$   &  $-1.931$   & $1/2,3/2$    \\ 
  QZ/SD23/10au            &   $151.1002$    &   $149.7506$    &   $       $  &  $149.3007$   &  $-2.249$   & $1/2,3/2$    \\ 
  QZ/SD33/10au            &   $151.7513$    &   $150.3923$    &   $147.6753$ &  $149.9393$   &  $-2.265$   & $1/2,3/2$    \\ 
  QZ/SD33/10au            &                 &                 &              &  $149.9395$   &  $-2.264$   & $5/2,3/2$    \\ 
 \end{tabular}
 \end{center}
\end{table}
%

In Ref. \cite{NatComm_Tm_2019} Eq. (5) is given
\begin{equation}
 \Delta\alpha_{\text{DC}}^s = \alpha^s_{5/2} - \alpha^s_{7/2} = -0.063(30) ~{\text{a.u.}}
 \label{EQ:NATCOMM2019_DA}
\end{equation}
from a combination of measurement and calculation. This being a negative quantity means that $\alpha^s_{7/2} > \alpha^s_{5/2}$.

The electronic ground state $^2F_{7/2}$ of the Tm atom can in the Hartree-Fock picture be represented by a $4f^{13}$ configuration 
written as $4f_{5/2}^6\, 4f_{7/2}^7$ in terms of Hartree-Fock spinors (the $j=5/2$ level is energetically lower than the
$j=7/2$ level). In turn, the first excited state $^2F_{5/2}$ can be represented as $4f_{5/2}^5\, 4f_{7/2}^8$. According to
Ref. \cite{desclaux_tables} numerical Dirac-Hartree-Fock calculations show that the radial expectation values of the valence
spinors are 
\begin{eqnarray*}
 \left<\hat{r}\right>_{5/2} &=& 0.763 ~{\text{a.u.}}  \\
 \left<\hat{r}\right>_{7/2} &=& 0.780 ~{\text{a.u.}}
\end{eqnarray*}
respectively. This means that, qualitatively, the level with the greater $4f_{7/2}$ occupation is in a straightforward
interpretation expected to be the level with the 
greater static dipole polarizability, since the $4f_{7/2}$ spinors are more diffuse than the $4f_{5/2}$ spinors. The level
with the greater $4f_{7/2}$ occupation is the $^2F_{5/2}$ level ($8$ electrons occupying the $j=7/2$ spinors). Therefore,
in Dirac-Hartree-Fock theory, the expected result is $\alpha^s_{5/2} > \alpha^s_{7/2}$ which contradicts the result given in
Ref. \cite{NatComm_Tm_2019} and in Eq. \ref{EQ:NATCOMM2019_DA} above.

Relativistic many-body calculations yield results for the differential scalar static dipole polarizability
defined as $\Delta\alpha_0 = \alpha_0(^2F_{5/2}) - \alpha_0(^2F_{7/2})$ and compiled in Table \ref{TAB:DELTA_ALPHA}.

\begin{table}[h]
 \begin{center}
   \caption{\label{TAB:DELTA_ALPHA} 
     Differential static scalar electric dipole polarizabilities $\alpha_0$ [a.u.] for the thulium atom ground term
     calculated through the FF method from various CI models
   }
  \vspace*{0.3cm}
  \begin{tabular}{l|cc|c}
   Model               &  $\alpha_0(^2F_{5/2})$  &  $\alpha_0(^2F_{7/2})$  &  $\Delta\alpha_0$   \\ \hline
   TZ/DCHF/0au         &     $184.6092$          &     $184.5674$          &     $ 0.0418$       \\
   TZ/SD15/6au         &     $162.3087$          &     $162.1495$          &     $ 0.1592$       \\
   TZ/SD13\_SD15/6au   &     $162.3028$          &     $162.1442$          &     $ 0.1586$       \\
   TZ/SD13\_SDT15/6au  &     $160.4280$          &     $160.3407$          &     $ 0.0873$       \\
   TZ/SD13\_SDTsppdQ15/6au & $165.4374$          &     $165.4144$          &     $ 0.0230$       \\ 
   QZ/DCHF/0au         &     $184.7382$          &     $184.6948$          &     $ 0.0434$       \\
   QZ/SD15/10au        &     $159.4748$          &     $159.3583$          &     $ 0.1165$       \\
   QZ/SDT15/10au       &     $159.2004$          &     $159.1952$          &     $ 0.0052$       \\
   QZ/SD23/10au        &     $149.3007$          &     $149.2431$          &     $ 0.0576$       \\
   QZ/SD33/10au        &     $149.9393$          &     $149.8976$          &     $ 0.0417$       \\ \hline
   {\bf{final}}        &                         &                         &     ${\boldsymbol{-0.134 \pm 0.11}}$   \\ \hline
   Experiment \cite{NatComm_Tm_2019}  &          &                         &     $-0.063(30)$
  \end{tabular}
 \end{center}
\end{table}

It is interesting to note that at all individual levels of calculation the qualitative picture of DCHF theory is reproduced
also when electron correlation effects are taken into account. However, when adding the effect of Triple excitations
to the model with the greatest number of electrons subjected to the correlation treatment, SD33/10au, the mentioned
interpretation at Hartree-Fock level of theory is reverted. Another sizable negative correction from a limited set of
quadruple excitations on top of the triple excitations is determined by using the smaller TZ basis set where a calculation
of this size becomes feasible. Going beyond this model is not possible with the current implementation due to computational
limitations and calculation time.

The final result for the differential static scalar dipole polarizability for the Tm atom $^2F_{7/2}(F=4)$--$^2F_{5/2}(F=3)$ 
clock transition is obtained by using the result from the model with the greatest number of electrons in the
correlation treatment (SD33/10au) as a base value and adding to it corrections due to CI excitation ranks surpassing
singles and doubles excitations. In formal terms this calculation reads as

\begin{eqnarray}
 \Delta\alpha_0({\text{final}}) &=& \Delta\alpha_0({\text{QZ/SD33/10au}})  \\
 \nonumber
             &&  + \Delta\alpha_0({\text{QZ/SDT15/10au}}) - \Delta\alpha_0({\text{QZ/SD15/10au}})  \\
 \nonumber
             &&  + \Delta\alpha_0({\text{TZ/SD13\_SDTsppdQ15/6au}}) - \Delta\alpha_0({\text{TZ/SD13\_SDT15/6au}})
\end{eqnarray}

Physically speaking, an encompassing treatment of interelectron correlation effects leads to the astonishing result that
the static scalar electric dipole polarizability is greater in the $^2F_{7/2}$ state than in the $^2F_{5/2}$ state of thulium. 
The simple single-determinant picture for electric polarizability based on Hartree-Fock spinors, therefore, breaks down.
The present result confirms qualitatively in this regard the result from Ref.
\cite{NatComm_Tm_2019} that has been obtained through a combination of experimental measurement and theory.
However, due to the sizable corrections found with even the most extensive CI models the uncertainty on the present result
must remain rather large. The uncertainty based on the limited treatment of CI excitation ranks, atomic basis sets, 
correlation effects from inner atomic shells and approximations in the employed atomic Hamiltonian does not compromise the 
qualitative conclusions.

\section{Conclusions}
\label{SEC:CONCL}
A variational relativistic configuration-interaction approach to the calculation of the hyperfine Stark coefficient 
in atoms is presented. 
As a methodological conclusion, the present approach can be applied to electronic transitions of any type in any atom, 
given that the hyperfine Stark coefficient is calculated as a differential static polarizability.

For the $^{87}$Rb atom excitation ranks up to quintuples have been included in the wavefunction expansion and the
obtained central value is the theoretical result so far closest to the experimental value. Calculations of similar
sophistication for the $^{133}$Cs atom yield a final result that is compatible with other high-level theoretical 
calculations and the 1998 experimental value by Simon {\it{et al.}} \cite{Simon_PRA1998}. 
The present method is then applied to a $^{169}$Tm clock transition where so far pure ab-initio calculations have
been lacking. The present calculations explain how the sign of the hyperfine Stark coefficient that has previously
been measured \cite{NatComm_Tm_2019} comes about. 
The difficulties in obtaining accurate electron correlation effects for relevant
properties of $^{169}$Tm have also been encountered in the calculation of its ground-state electric quadrupole
moment \cite{Fleig_PhysRevA.107.032816}. 

\clearpage
\bibliographystyle{unsrt}
\newcommand{\Aa}[0]{Aa}

\clearpage

\end{document}